\documentclass[twocolumn]{aastex62}
\usepackage{newtxtext,newtxmath}
\usepackage[T1]{fontenc}
\usepackage{ae,aecompl}
\usepackage{graphicx}
\usepackage{amsmath}
\usepackage{amssymb}
\usepackage{footnote}
\usepackage{epstopdf}

\def\wid{WASP-190}
\def\widb{WASP-190b}

\def\teff{$T_{\rm eff}$}
\def\logg{$\log{g_{*}}$}
\def\vsini{$v \sin i_{\star}$}

\submitjournal{AJ}

\shorttitle{WASP-190b}
\shortauthors{Temple et al.}

\begin{document}

\title{WASP-190b: Tomographic discovery of a transiting hot Jupiter.}

\correspondingauthor{L. Y. Temple}
\email{l.y.temple@keele.ac.uk}

\author{L. Y. Temple}
\affil{Astrophysics Group, Keele University, Staffordshire, ST5 5BG, UK}

\author{C. Hellier}
\affil{Astrophysics Group, Keele University, Staffordshire, ST5 5BG, UK}

\author{Y. Almleaky}
\affil{Space and Astronomy Department, Faculty of Science, King Abdulaziz University, \\
21589 Jeddah, Saudi Arabia}
\affil{King Abdullah Centre for Crescent Observations and Astronomy (KACCOA), Makkah Clock, Saudia Arabia}

\author{D.R. Anderson}
\affil{Astrophysics Group, Keele University, Staffordshire, ST5 5BG, UK}

\author{F. Bouchy}
\affil{Observatoire astronomique de l'Universit\'e de Gen\`eve \\
51 ch. des Maillettes, 1290 Sauverny, Switzerland}

\author{D.J.A. Brown}
\affil{Department of Physics, University of Warwick, \\
Gibbet Hill Road, Coventry, CV4 7AL, UK}
\affil{Centre for Exoplanets and Habitability, University of Warwick, \\
Gibbet Hill Road, Coventry CV4 7AL, UK}

\author{A. Burdanov}
\affil{Space sciences, Technologies and Astrophysics Research (STAR) Institute, \\
Universit{\'e} de Li{\`e}ge, All{\'e}e du 6 Ao{\^u}t 17, 4000 Li{\`e}ge, Belgium}

\author{A. Collier Cameron}
\affil{SUPA, School of Physics and Astronomy, University of St.\ Andrews \\
North Haugh,  Fife, KY16 9SS, UK }

\author{L. Delrez}
\affil{Cavendish Laboratory, J J Thomson Avenue, \\
Cambridge, CB3 0HE, UK}

\author{M. Gillon}
\affil{Space sciences, Technologies and Astrophysics Research (STAR) Institute, \\
Universit{\'e} de Li{\`e}ge, All{\'e}e du 6 Ao{\^u}t 17, 4000 Li{\`e}ge, Belgium}


\author{E. Jehin}
\affil{Space sciences, Technologies and Astrophysics Research (STAR) Institute, \\
Universit{\'e} de Li{\`e}ge, All{\'e}e du 6 Ao{\^u}t 17, 4000 Li{\`e}ge, Belgium}

\author{M. Lendl}
\affil{Space Research Institute, Austrian Academy of Sciences, Schmiedlstr. 6, 80 42, Graz, Austria}
\affil{Observatoire astronomique de l'Universit\'e de Gen\`eve \\
51 ch. des Maillettes, 1290 Sauverny, Switzerland}

\author{P.F.L. Maxted}
\affil{Astrophysics Group, Keele University, Staffordshire, ST5 5BG, UK}

\author{C. Murray}
\affil{Cavendish Laboratory, J J Thomson Avenue, Cambridge, CB3 0HE, UK}

\author{L. D. Nielsen}
\affil{Observatoire astronomique de l'Universit\'e de Gen\`eve \\
51 ch. des Maillettes, 1290 Sauverny, Switzerland}

\author{F. Pepe}
\affil{Observatoire astronomique de l'Universit\'e de Gen\`eve \\
51 ch. des Maillettes, 1290 Sauverny, Switzerland}

\author{D. Pollacco}
\affil{Department of Physics, University of Warwick, \\
Gibbet Hill Road, Coventry, CV4 7AL, UK}
\affil{Centre for Exoplanets and Habitability, University of Warwick, \\
Gibbet Hill Road, Coventry CV4 7AL, UK}

\author{D. Queloz}
\affil{Cavendish Laboratory, J J Thomson Avenue, Cambridge, CB3 0HE, UK}

\author{D. S\'egransan}
\affil{Observatoire astronomique de l'Universit\'e de Gen\`eve \\
51 ch. des Maillettes, 1290 Sauverny, Switzerland}

\author{B. Smalley}
\affil{Astrophysics Group, Keele University, Staffordshire, ST5 5BG, UK}


\author{S. Thompson}
\affil{Cavendish Laboratory, J J Thomson Avenue, Cambridge, CB3 0HE, UK}

\author{A.H.M.J. Triaud}
\affil{School of Physics \& Astronomy, University of Birmingham, Edgbaston, Birmingham, B15 2TT, UK}

\author{O.D. Turner}
\affil{Observatoire astronomique de l'Universit\'e de Gen\`eve \\
51 ch. des Maillettes, 1290 Sauverny, Switzerland}
\affil{Astrophysics Group, Keele University, Staffordshire, ST5 5BG, UK}

\author{S. Udry}
\affil{Observatoire astronomique de l'Universit\'e de Gen\`eve \\
51 ch. des Maillettes, 1290 Sauverny, Switzerland}

\author{R.G. West}
\affil{Department of Physics, University of Warwick, \\
Gibbet Hill Road, Coventry, CV4 7AL, UK}

\begin{abstract}

We report the discovery of WASP-190b, an exoplanet on a 5.37-day orbit around a mildly-evolved F6 IV-V star with $V$\,=\,11.7, \teff\,=\,6400 $\pm$ 100\,K, $M_{\rm *}$\,=\,1.35\,$\pm$\,0.05\,$M_{\odot}$ and $R_{\rm *}$\,=\,1.6\,$\pm$\,0.1\,$R_{\odot}$. The planet has a radius of $R_{\rm P}$\,=\,1.15\,$\pm$\,0.09\,$R_{\rm Jup}$ and a mass of $M_{\rm P}$\,=\,1.0\,$\pm$\,0.1\,$M_{\rm Jup}$, making it a mildly inflated hot Jupiter. It is the first hot Jupiter confirmed via Doppler tomography with an orbital period >\,5\,days. The orbit is also marginally misaligned with respect to the stellar rotation, with $\lambda$\,=\,21\,$\pm$\,6$^{\circ}$ measured using Doppler tomography. 

\end{abstract}


\section{Introduction} \label{sec:intro}
The Rossiter--McLaughlin (RM) effect, a distortion of the line profiles of a star caused by an occulting body blocking part of the stellar face, was first detected for a transiting hot Jupiter by \citet{2000A&A...359L..13Q} in observations of HD\,209458, whereby the distortion was detected as a perturbation to radial velocity measurements. It has since been used extensively to measure the projected angle between the planet's orbit and the stellar rotation axis in many hot-Jupiter systems \citep[e.g.][]{2017haex.bookE...2T}. The current alignment of a planetary orbit with respect to the stellar rotation is an indicator of the dynamical history of the planet, and can point to the formation mechanisms at play. 

One can also plot the line profiles as a function of phase, looking for the Doppler shadow of the planet as it moves across the line profiles. The detection of the Doppler tomographic signal of a candidate exoplanet can rule out transit mimics such as blended eclipsing binaries. This ``tomographic'' method was first used in the discovery of a planet for WASP-33b \citep{2010MNRAS.407..507C}. The tomographic technique is particularly useful for systems with hotter and fast-rotating stars, with fewer and broader spectral lines, which may give only less-accurate radial-velocity measurements  and thus were previously paid less attention by transit and radial velocity surveys.  Thus Doppler tomography has now been used in the discovery of hot Jupiters transiting hot stars, including KELT-20b \citep{2017AJ....154..194L}, HAT-P-67b \citep{2017AJ....153..211Z}, WASP-167b/KELT-13b \citep{2017MNRAS.471.2743T}, MASCARA-1b \citep{2017A&A...606A..73T} and WASP-189b \citep{2018arXiv180904897A}. 

\citet{2017MNRAS.464..810B} compare tomographic and RM analyses of the same datasets for six WASP systems.  They find that the tomographic method consistently gives better constraints on values for the projected stellar rotational velocity \vsini\ and the sky-projected obliquity angle $\lambda$.  Note that the tomographic analysis method uses the line profiles more directly, while an RM analysis in terms of radial-velocity measurements (RVs) needs one to translate the change in the line profiles owing to a planet shadow into a change in the overall radial velocity \citep[e.g.][]{2011ApJ...742...69H,2013A&A...550A..53B}. 

The number of known exoplanets has grown to the point where population studies can draw significant and meaningful conclusions about their bulk properties and dynamical histories. For example, \citet{2018ApJ...853...37S} uses a sample of 146 systems comprised of a solar-like star and a giant planet, brown dwarf, or low-mass stellar companion, to place a mass-limit boundary between hot Jupiters and brown dwarfs which relies on their formation mechanism. It is not possible, however, to perform the same scale of population studies for hot Jupiters orbiting early-type stars, due to the relative lack of such objects discovered so far (resulting, at least partially, from selection biases in past transit and radial-velocity surveys).

Hot Jupiters orbiting hot stars are of interest due to the orbital and physical differences between them and hot Jupiters orbiting later-type stars. They are more likely to be in misaligned orbits \citet{2010ApJ...718L.145W,2017haex.bookE...2T}, often have stars which rotate more quickly than the planet's orbit \citep[e.g.][]{2017AJ....153...94C,2017MNRAS.471.2743T}, and are inflated, with hotter dayside temperatures, due to the increased irradiation from their host star \citep{2016AJ....152..182H,2018MNRAS.480.5307T}. The increased irradiation might also result in planetary magnetic fields which are stronger than in cooler Jupiters, since the continuous injection of energy into the interior of a gas giant might produce a more efficient planetary dynamo \citep{2017ApJ...849L..12Y}.

In this work, we report the discovery and characterisation of \widb, a hot Jupiter orbiting a star of \teff\,=\,6400\,K which can be found in {\it TESS} Sector 2 as TIC ID:116156517 \citep{2015JATIS...1a4003R,2018AJ....156..102S}. We use both tomographic and RM analyses to determine the geometry of the system, and confirm the existence of the planet via the detection of its Doppler shadow and by measuring its mass using orbital RV measurements.

\section{Data and observations} \label{sec:obs}
\begin{table}
\caption{Observations of \widb.}
\centering
\begin{tabular}{lcc}
\hline
Telescope/Instrument & Date & Notes \\[0.5ex]
\hline
WASP-South & 2006--2011 & 30137 pts. \\
TRAPPIST-South & 2014 Nov 26 & {\it I+z}, 7s exp. \\
SPECULOOS-Europa & 2017 Oct 13 & {\it I+z}, 10s exp. \\
CORALIE & 2014 Aug--Oct & 5 RVs \\
HARPS & 2017 Oct 13 & 28 spectra taken \\
 & & including a transit \\
HARPS & 2018 Oct & 5 RVs \\
\hline
\end{tabular}
\label{table:observations}
\end{table}

\begin{table}
\caption{RV measurements of \wid, taken using the CORALIE and HARPS spectrographs for this work.}
\centering
\begin{tabular}{lccrc}
\hline
BJD$_{TDB}$ & RV & $\sigma$$_{\rm RV}$ & BS & $\sigma$$_{\rm BS}$ \\
--2,450,000  & (km s$^{-1}$) & (km s$^{-1}$) & (km s$^{-1}$) & (km s$^{-1}$) \\ [0.5mm]
\hline
\multicolumn{5}{l}{CORALIE (out of transit):} \\
6871.794771	& 0.89 & 0.05 & --0.07 & 0.10 \\
6895.811527	& 0.94 & 0.05 & --0.16 & 0.10 \\
6922.731329	& 0.90 & 0.03 & --0.05 & 0.06 \\
6952.511870	& 0.95 & 0.04 & 0.15 & 0.08 \\
7000.625106 & 0.87 & 0.04 & --0.22 & 0.08 \\
8392.595422	\medskip & 0.84 & 0.07 & --0.24 & 0.14\\
\multicolumn{5}{l}{HARPS (including a transit):} \\
8040.529251 & 0.84 & 0.02 & 0.04 & 0.04 \\
8040.540026 & 0.82 & 0.02 & --0.04 & 0.04 \\
8040.550489 & 0.84 & 0.02 & 0.05 & 0.04 \\
8040.561472 & 0.86 & 0.02 & --0.02 & 0.04 \\
8040.572351 & 0.85 & 0.02 & --0.05 & 0.04 \\
8040.582918 & 0.85 & 0.02 & --0.02 & 0.04 \\
8040.593797 & 0.87 & 0.02 & --0.00 & 0.04 \\
8040.604572 & 0.92 & 0.02 & 0.02 & 0.04 \\
8040.615243 & 0.89 & 0.02 & --0.07 & 0.04 \\
8040.626122 & 0.90 & 0.02 & --0.13 & 0.04 \\
8040.636897 & 0.87 & 0.02 & --0.06 & 0.04 \\
8040.647776 & 0.84 & 0.02 & --0.01 & 0.04 \\
8040.658459 & 0.80 & 0.02 & --0.02 & 0.04 \\
8040.669130 & 0.79 & 0.02 & 0.04 & 0.04 \\
8040.680113 & 0.80 & 0.02 & 0.10 & 0.04 \\
8040.690784 & 0.80 & 0.02 & 0.03 & 0.04 \\
8040.701350 & 0.78 & 0.02 & --0.07 & 0.04 \\
8040.712334 & 0.76 & 0.02 & --0.03 & 0.04 \\
8040.723016 & 0.73 & 0.02 & --0.03 & 0.04 \\
8040.733791 & 0.76 & 0.02 & --0.05 & 0.04 \\
8040.744670 & 0.73 & 0.03 & --0.05 & 0.06 \\
8040.755341 & 0.76 & 0.03 & 0.05 & 0.06 \\
8040.766220 & 0.77 & 0.03 & 0.09 & 0.06 \\
8040.777100 & 0.81 & 0.03 & --0.10 & 0.06 \\
8040.787770 & 0.83 & 0.03 & --0.02 & 0.06 \\
8040.798545 & 0.82 & 0.03 & 0.10 & 0.06 \\
8040.809112 & 0.85 & 0.03 & --0.06 & 0.06 \\
8040.820107 \medskip & 0.82 & 0.03 & 0.06 & 0.06 \\
\multicolumn{5}{l}{HARPS (out of transit):} \\
8393.843700 & 0.92 & 0.01 & --0.20 & 0.02 \\
8396.706300 & 0.73 & 0.01 & 0.10 & 0.02 \\
8397.590800 & 0.86 & 0.04 & --0.15 & 0.08 \\
8398.611000 & 0.92 & 0.02 & --0.24 & 0.04 \\
8399.542350 & 0.85 & 0.04 & --0.13 & 0.08 \\
\hline
\end{tabular}
\label{table:RVs}
\end{table}

We observed \wid\ using the WASP-South telescope \citep{2011EPJWC..1101004H} at the South African Astronomical Observatory (SAAO) from 2006 to 2011. After the detection of a planet-like transit dip in the WASP lightcurve we confirmed the transit with a  follow-up lightcurve obtained using the TRAPPIST-South telescope \citep{2011Msngr.145....2J}, and proceeded to obtain reconnaissance spectroscopy with the Euler/CORALIE spectrograph \citep{2001Msngr.105....1Q}. These were sufficient to rule out a stellar-mass binary, but with relatively large errors were consistent with no motion at the level of 250\,m\,s$^{-1}$ and were inconclusive about whether the transiting body was a planet. 

We thus attempted tomography of a transit, obtaining a series of 28 spectra through transit on the night of 2017 October 13 using the ESO 3.6-m/HARPS spectrograph \citep{2002Msngr.110....9P}, accompanied by simultaneous photometry using the SPECULOOS-Europa telescope \citep{2017haex.bookE.130B,2018NatAs...2..344G,2018SPIE10700E..1ID}. After tomographic detection of a planet-like signal, we obtained 5 further orbital RVs with HARPS to constrain the planetary mass.

The HARPS spectra were cross-correlated over a window of $\pm$\,350\,km\,s$^{-1}$, using a mask matching a G2 spectral type, and the standard HARPS Data Reduction Software as described by \citet{1996A&AS..119..373B}, \citet{2002Msngr.110....9P}.  We then analysed the cross-correlation functions (CCFs) themselves, and computed radial velocity (RV) measurements from the CCFs which we list in Table~\ref{table:RVs} along with bisector spans (BS).

We used the WASP photometric data to look for any evidence of rotational modulation of the host star, using the methods of \citet{2011PASP..123..547M}. We find no such variability at periods longer than a day, with a 95\%-confidence upper limit on the amplitude of 1\,mmag.

\section{Stellar parameters from spectral analysis} \label{sec:specanalysis}
In order to determine stellar parameters of \wid\ we co-added the HARPS spectra obtained on the night of 2017 Oct 13 and performed a spectral analysis. We adopted a microturbulent velocity of $v_{\rm mic}$\,=\,1.6\,km\,s$^{-1}$ from the calibration of \citet{2010MNRAS.405.1907B} and a macroturbulent velocity of $v_{\rm mac}$\,=\,6.5\,km\,s$^{-1}$ from the calibration of \citet{2014MNRAS.444.3592D}. We used the H$\alpha$ line to determine an effective temperature \teff\,=\,6400\,$\pm$\,100\,K, while using the Na D feature to measure \logg\,=\,3.9\,$\pm$\,0.1. We also determined the projected stellar rotational velocity \vsini\,=\,13.8\,$\pm$\,0.7\,km\,s$^{-1}$, and the surface metallicity [Fe/H]\,=\,$-$0.02\,$\pm$\,0.05. These results are also listed in Table~\ref{table:allResults}.   Using the MKCLASS program \citep{2014AJ....147...80G} we then obtained a spectral type of F6 IV--V. 

\begin{figure}
\hspace*{2mm}\includegraphics[width=0.49\textwidth]{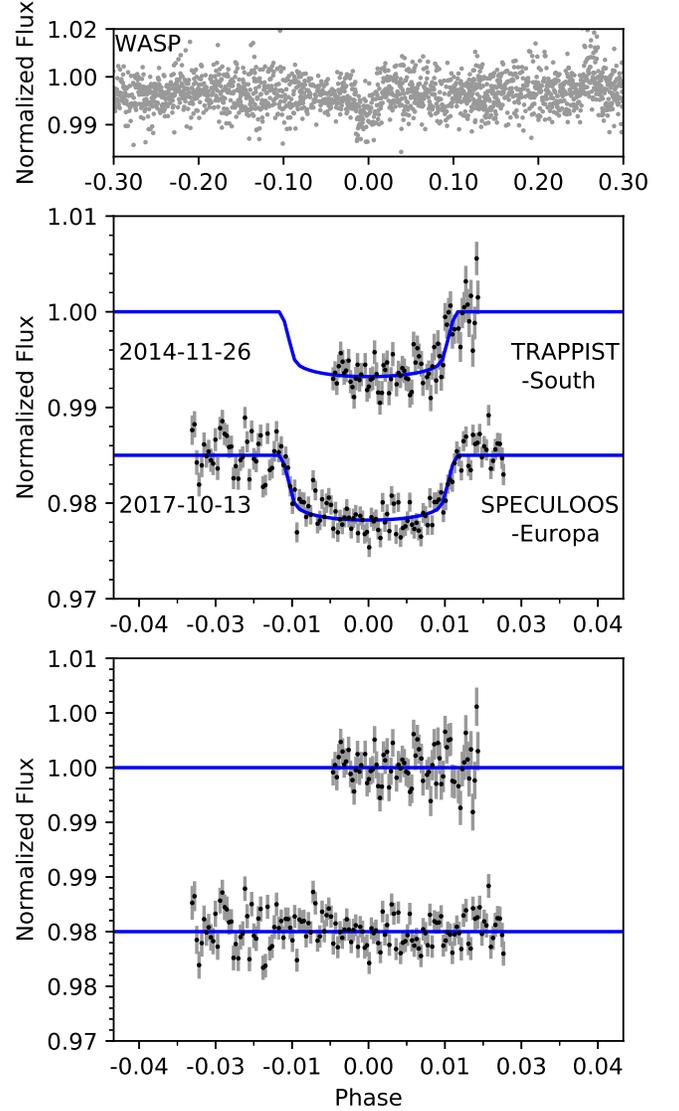}\\ [-2mm]
\caption{Top: the discovery lightcurve for \widb\ (WASP-South). Middle: the two follow-up lightcurves with the best-fitting model shown in blue. Bottom: the residuals of the fits to the follow-up lightcurves.}
\label{fig:phot}
\end{figure}

\section{Combined MCMC analysis}
\label{sec:analysis}
\begin{table}
\caption{All system parameters obtained in the combined analyses for \widb\ .} 
\label{table:allResults}
\centering
\begin{tabular}{lcc}
\hline
\multicolumn{3}{l}{1SWASP\,J003050.23--403424.3}\\
\multicolumn{3}{l}{2MASS\,00305023--4034243}\\
\multicolumn{3}{l}{TIC ID:116156517}\\
\multicolumn{3}{l}{Gaia DR2 4994237247949280000}\\
\multicolumn{3}{l}{RA\,=\,00$^{\rm h}$30$^{\rm m}$50.233$^{\rm s}$, 
Dec\,=\,--40$^{\circ}$34$^{'}$24.36$^{''}$ (J2000)}\\
\multicolumn{3}{l}{$V$ = 11.7\,$\pm$\,0.1 (TYCHO2)}  \\
\multicolumn{3}{l}{{\it Gaia} DR2 Proper Motions:} \\
\multicolumn{3}{l}{(RA) 38.23\,$\pm$\,0.03 (Dec) --9.14\,$\pm$\,0.04 mas/yr} \\
\multicolumn{3}{l}{{\it Gaia} DR2 Parallax: 1.82\,$\pm$\,0.03\,mas} \\
\multicolumn{3}{l}{Rotational Modulations: < 1 mmag (95\%)}\\
\hline
\multicolumn{3}{l}{\it{Stellar parameters from spectral analysis:}} \\[0.5ex]
Parameter & \multicolumn{2}{c}{Value} \\
(Unit) & & \\
Spectral type & \multicolumn{2}{c}{F6 IV--V} \\
\teff\ (K) & \multicolumn{2}{c}{6400\,$\pm$\,100} \\
$\log g_{*}$ & \multicolumn{2}{c}{3.9\,$\pm$\,0.1} \\
{[Fe/H]} & \multicolumn{2}{c}{--0.02\,$\pm$\,0.05} \\
$v \sin i_{\rm *}$ (km\,s$^{-1}$) & \multicolumn{2}{c}{$13.8\,\pm\,0.7$} \\
$v_{\rm mic}$ (km\,s$^{-1}$) & \multicolumn{2}{c}{1.6 (assumed)} \\
$v_{\rm mac}$ (km\,s$^{-1}$) \medskip & \multicolumn{2}{c}{6.5 (assumed)} \\
\multicolumn{3}{l}{\it{Parameters from photometric and RV analysis:}} \\[0.5ex]
Parameter & \multicolumn{2}{c}{DT Value} \\ 
(Unit) & \multicolumn{2}{c}{(adopted):} \\
$P$ (d) & \multicolumn{2}{c}{5.367753 $\pm$ 0.000004} \\ 
$T_{\rm c}$ (BJD$_{\rm TDB}$) & \multicolumn{2}{c}{2457799.1256 $\pm$ 0.0007} \\ 
$T_{\rm 14}$ (d) & \multicolumn{2}{c}{0.186 $\pm$ 0.002} \\ 
$T_{\rm 12}=T_{\rm 34}$ (d) & \multicolumn{2}{c}{0.017 $\pm$ 0.002} \\ 
$R_{\rm P}^{2}$/R$_{*}^{2}$ & \multicolumn{2}{c}{0.0062 $\pm$ 0.0002} \\ 
$b$ & \multicolumn{2}{c}{0.45 $\pm$ 0.09} \\ 
$i$ ($^\circ$) & \multicolumn{2}{c}{87.1 $\pm$ 0.7} \\ 
$a$ (AU)  & \multicolumn{2}{c}{0.0663 $\pm$ 0.0008} \\ 
$M_{\rm *}$ ($M_{\rm \odot}$) & \multicolumn{2}{c}{1.35 $\pm$ 0.05} \\ 
$R_{\rm *}$ ($R_{\rm \odot}$) & \multicolumn{2}{c}{1.6 $\pm$ 0.1} \\ 
$\log g_{*}$ (cgs) & \multicolumn{2}{c}{4.17 $\pm$ 0.04} \\ 
$\rho_{\rm *}$ ($\rho_{\rm \odot}$) & \multicolumn{2}{c}{0.34 $\pm$ 0.05} \\ 
$T_{\rm eff}$ (K) & \multicolumn{2}{c}{6400 $\pm$ 100} \\ 
{[Fe/H]} & \multicolumn{2}{c}{--0.02 $\pm$ 0.05} \\ 
$K$ (km s$^{-1}$) & \multicolumn{2}{c}{0.099 $\pm$ 0.009} \\ 
$M_{\rm P}$ ($M_{\rm Jup}$) & \multicolumn{2}{c}{1.0 $\pm$ 0.1} \\ 
$R_{\rm P}$ ($R_{\rm Jup}$) & \multicolumn{2}{c}{1.15 $\pm$ 0.09} \\ 
$\log g_{\rm P}$ (cgs) & \multicolumn{2}{c}{3.2 $\pm$ 0.1} \\ 
$T_{\rm eql}$ (K) \medskip & \multicolumn{2}{c}{1500 $\pm$ 50} \\ 
\multicolumn{3}{l}{\it{Parameters from RM and DT analyses:}}\\[0.5ex]
Parameter & DT Value & RM Value: \\
(Unit) & (adopted): & \\
$\gamma$ (km s$^{-1}$)& 0.82 $\pm$ 0.01 & 0.823 $\pm$ 0.009 \\
$\lambda$ ($^\circ$) & 21 $\pm$ 6 & 23 $\pm$ 12 \\
v$_{\rm FWHM}$ (km\,s$^{-1}$) & 10 $\pm$ 1 & -- \\
\vsini\ (km\,s$^{-1}$) \medskip & 13.3 $\pm$ 0.6 & 14.1 $\pm$ 0.7 \\
\hline
\end{tabular} 
\end{table}

We conduct an analysis very similar to that conducted by \citet{2018MNRAS.480.5307T} for WASP-174b, which involves the use of Markov Chain Monte Carlo (MCMC) methods to analyse the combined photometric and spectroscopic datasets. As one approach we use the in-transit spectroscopy data in the form of RV measurements, following the method of \citet{2011ApJ...742...69H},  and as a second approach we use the same data in the form of CCFs, following methods similar to that used by \citet{2017MNRAS.464..810B,2017MNRAS.471.2743T}. We call the former the Rossiter--McLaughlin (RM) analysis and the latter the tomographic analysis. 

The code we use is described by \citet{2007MNRAS.380.1230C} and \citet{2008MNRAS.385.1576P}, which in the latest version includes the tomographic analysis as described by \citet{2010MNRAS.403..151C}. In both analyses, fitting the photometric lightcurves allows direct measurement of the planet-to-star area ratio $(R_{\rm p}/R_{\star})^{2}$, the impact parameter $b$ and the key transit timing information $T_{\rm c}$, $P$, $T_{\rm 14}$ and by extension $T_{\rm 12}$, which are respectively the epoch of mid-transit, the orbital period, the transit duration and the duration of ingress (or equivalently egress). We use the value of \teff\ obtained in the spectral analysis (see discussion in Section~\ref{sec:discussion}) as the starting value for the MCMC chains, and for each new value of \teff\ we interpolate four-parameter law limb-darkening coefficients from the tables of \citet{2000A&A...363.1081C, 2004A&A...428.1001C}. Stellar mass is calculated at each step using the Enoch--Torres relation \citep{2010A&A...516A..33E, 2010A&ARv..18...67T}. The photometric data are displayed in Fig.~\ref{fig:phot} along with the best-fit model and residuals of the fit.

The RV analysis then enables measurement of the barycentric system velocity $\gamma$ and the stellar reflex velocity semi-amplitude $K$. We expect that most hot Jupiters will settle into a circular orbit on a shorter timescale than their lifetimes \citep{2011MNRAS.414.1278P}, but with an orbital period of $\sim$\,5\,days \widb\ is entering the regime where eccentricity may remain. However, we do not have sufficient orbital RVs to constrain the eccentricity and so assume a circular orbit. We do not include the CORALIE measurements in the model adopted here, although including them changes the planetary mass by much less than the error bar.

Both the RM analysis and the tomographic analysis can allow the measurement of \vsini\ and $\lambda$, while providing an additional constraint on the values of $\gamma$ and $b$. However, it can often be the case that for RM analysis a prior on \vsini\ is required in order to obtain a well-constrained fit, and so we adopt the spectral \vsini\ as a prior for both analyses. In the tomographic analysis we also fit the local line width $v_{\rm FWHM}$, resulting from stellar turbulence and instrumental broadening, which influences the width of the planetary perturbation of the line profiles, and whose shape is assumed to be Gaussian.

\begin{figure}
\hspace*{2mm}\includegraphics[width=0.47\textwidth]{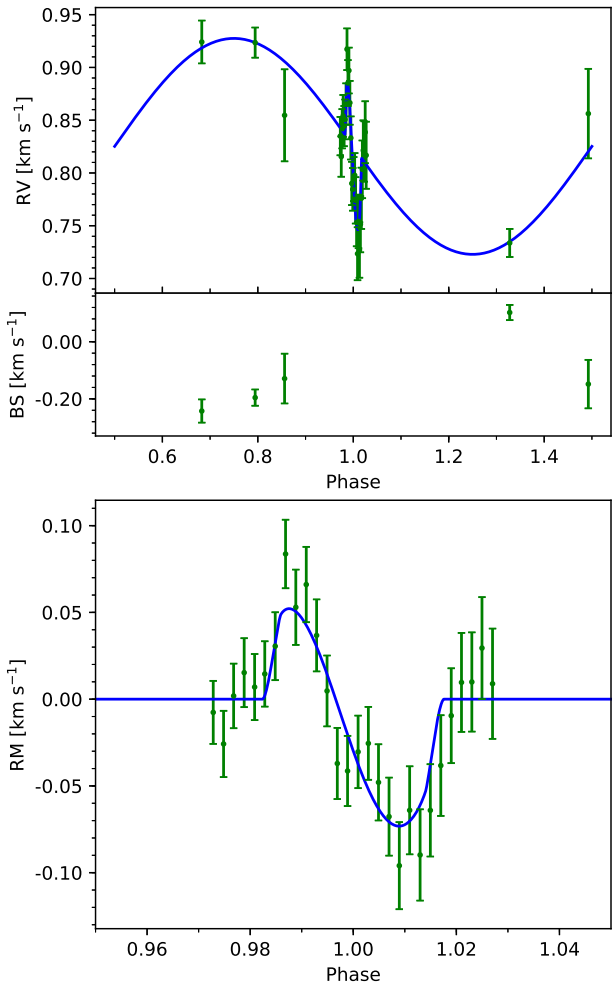}\\ [-2mm]
\caption{Top: The HARPS RV measurements used in the analysis of \widb. The blue line shows the best-fit Keplerian RV curve and the fit to the RM effect. Centre: the bisectors for the out-of-transit RVs plotted against phase, which show no correlation with the RV measurements. Bottom: The region around transit on a larger scale.}
\label{fig:RVRM}
\end{figure}

\begin{figure*}
\hspace*{2mm}\includegraphics[width=0.99\textwidth]{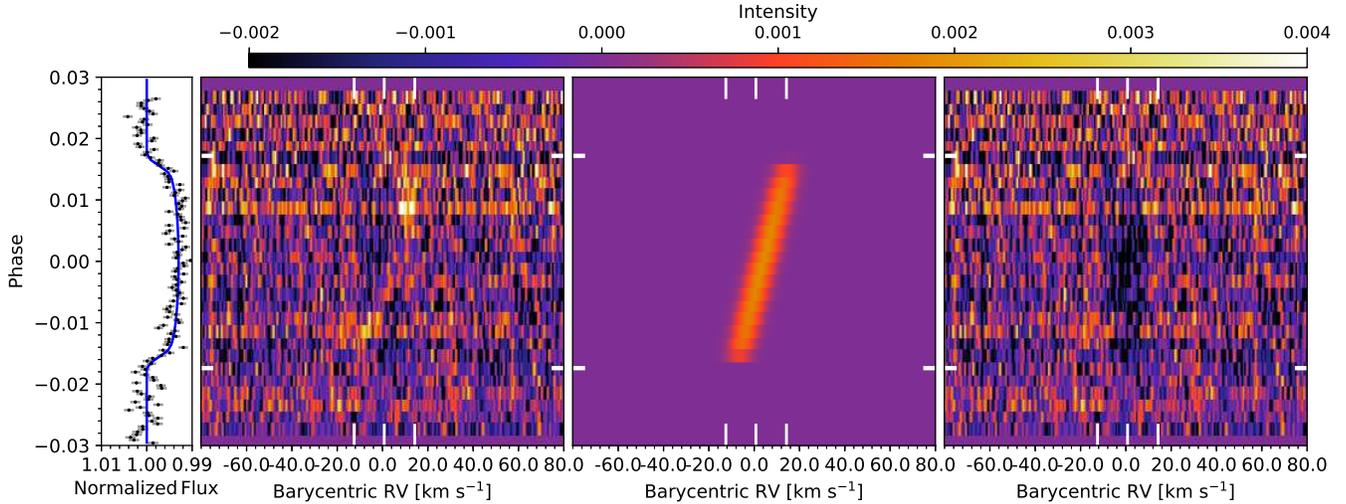}\\ [-2mm]
\caption{Centre-left: The Doppler tomogram comprised of the time series of residual HARPS CCFs calculated by subtracting the average of the out-of-transit CCFs from all CCFs. Left: the SPECULOOS-Europa lightcurve taken simultaneously with the HARPS observation. Centre-right: The best-fit planet model. Right: the residuals remaining after subtracting the best-fit planet model from the centre-left tomogram. In the three tomographic panels, the start and end times of the transit are marked with horizontal white dashes, while the vertical dashes mark respectively the positions of $\gamma$-\vsini\ , $\gamma$ and $\gamma$+\vsini\ . We interpret the tomogram as showing a very faint, prograde planet signal which in places is completely masked by background noise.}
\label{fig:tomog}
\end{figure*}

We show all RV measurements used in this work, along with the best fitting RV and RM models in Fig.~\ref{fig:RVRM}. We also display the tomographic data (the time series of CCFs with the average of the out-of-transit CCFs subtracted from all CCFs) in Fig.~\ref{fig:tomog}, along with the best-fit planet model and residuals. The best fit parameters are listed in Table~\ref{table:allResults}. We adopt the solution to the tomographic analysis (see Section~\ref{sec:planetresults}) and, to avoid duplicating parameters derived from the same data (which are consistent in any case), the only parameters for which we list values from both analyses are $\gamma$, \vsini\ and $\lambda$.

\section{Results for the star}
\label{sec:starresults}
\begin{figure}
\hspace*{2mm}\includegraphics[width=0.47\textwidth]{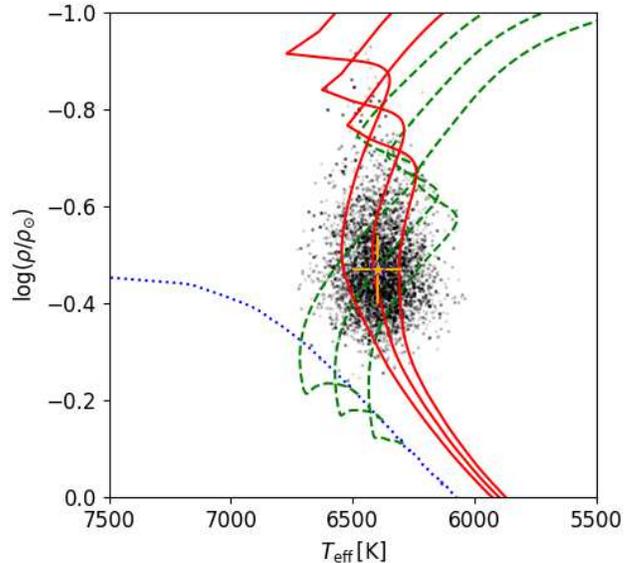}\\ [-2mm]
\caption{The best fitting evolutionary tracks and isochrones of \wid\ obtained using {\sc bagemass}. Black points: individual realisations of the MCMC. Dotted blue line: Zero-Age Main Sequence (ZAMS) at best-fit [Fe/H]. Green dashed lines: evolutionary track for the best-fit [Fe/H] and mass, plus $1\sigma$ bounds. Red lines: isochrone for the best-fit [Fe/H] and age, plus $1\sigma$ bounds. Orange star: measured values of \teff\ and $\rho_{*}$ for \wid\ obtained in the spectral and photometric analyses respectively.}
\label{fig:bagemass}
\end{figure}

\begin{table}
\caption{Parameters for \wid\ from {\sc bagemass}:} 
\label{table:bagemass}
\centering
\begin{tabular}{lc}
\hline
Parameter (Unit) & Value \\
\hline
Age (Gyr) & 2.8\,$\pm$\,0.4 \\
$M_{*}$ ($M_{\rm \odot}$) & 1.30\,$\pm$\,0.05 \\
$\rm{[Fe/H]}_{\rm init}$ & 0.03\,$\pm$\,0.04 \\
\hline
\end{tabular} 
\end{table}

We find \wid\ to have a large radius of $R_{\rm *}$\,= 1.6\,$\pm$\,0.1\,$R_{\rm \odot}$ and a density of $\rho_{\rm *}$\,= 0.34\,$\pm$\,0.05\,$\rho_{\rm \odot}$. This implies that the star is beginning to evolve away from the main sequence, which would be consistent with the spectral type of F6 IV--V.

The effective temperature ($T_{\rm eff}$) was also obtained using the Infrared Flux Method \citep[IRFM][]{1977MNRAS.180..177B}. The stellar spectral energy distribution (SED) was obtained using literature broad-band photometry from 2MASS \citep{2006AJ....131.1163S}, APASS9 $B$, $V$, $g'$, $r'$ and $i'$ \citep{2015AAS...22533616H}, USNO-B1 $R$ \citep{2003AJ....125..984M} and WISE \citep{2012yCat.2311....0C}. The photometry was converted to fluxes and the best-fitting \citet{1993KurCD..13.....K} model flux distribution found and
integrated to determine a bolometric flux of $5.27 \pm 0.26 \times 10^{-10}$~erg\,s$^{-1}$\,cm$^{-2}$. No visible interstellar lines were seen around the Na D line, so $E(B-V)$ was assumed to be zero. The IRFM was then used, with the 2MASS fluxes, to obtain a value of $T_{\rm eff} = 6560 \pm 140$~K as well as an angular diameter of $\theta$\,=\,0.029\,$\pm$\,0.001\,mas. The {\it Gaia} DR2 \citep{2016A&A...595A...1G,2018A&A...616A...1G} lists the parallax of \wid\ as 1.82\,$\pm$\,0.03\,mas. Using these values and accounting for the correction to {\it Gaia} DR2 parallax values suggested by \citet{2018ApJ...862...61S}, we obtain a stellar radius of 1.65\,$\pm$\,0.08\,$R_{\odot}$, which is consistent with our result from the MCMC analysis. 

We investigate the age of \wid\ using the open source software {\sc bagemass}\footnote{\url{http://sourceforge.net/projects/bagemass}} \citep{2015A&A...575A..36M}. {\sc bagemass} allows the user to fit \teff\ and $M_{\rm *}$ using stellar evolutionary models calculated for different He abundances and mixing lengths \citep[{\sc garstec};][]{2008Ap&SS.316...99W}. As inputs we use the values of \teff\ and [Fe/H] derived from the spectral analysis in Section~\ref{sec:specanalysis}, and also use the value of $\rho_{\rm *}$ obtained in the combined analysis (Section~\ref{sec:analysis}) as a constraint.

Assuming solar values for the He abundance and mixing length gave the best-fit solution. We display the corresponding isochrones and evolutionary tracks in Fig.~\ref{fig:bagemass}. We find the current age of \wid\ to be 2.8\,$\pm$\,0.4\,Gyr, implying that the star is beginning to evolve off the main sequence. This is consistent with our finding that the star has a radius larger than expected for a main sequence star. For comparison, the time taken to exhaust all hydrogen in the core is 3.8\,$\pm$\,0.5\,Gyr.

\section{Results for the planet}
\label{sec:planetresults}
We find a best fit $K$ of 0.099\,$\pm$\,0.009\,km\,s$^{-1}$, giving a planet mass of $M_{\rm p}$\,=\,1.0\,$\pm$\,0.1\,$M_{\rm Jup}$. The fitted planetary radius is  1.15\,$\pm$\,0.09\,R$_{\rm Jup}$. 

The in-transit RVs, showing the RM effect, are displayed in the lower panel of Fig.~\ref{fig:RVRM}. The equivalent tomogram of the same data is shown in Fig.~\ref{fig:tomog}. Both are consistent with a planet in a prograde orbit.  The projected spin-orbit angle, $\lambda$, is measured as 23\,$\pm$\,12$^{\circ}$ in the RM analysis and as 21\,$\pm$\,6$^{\circ}$ in the tomographic analysis. The planet trace is faint and hard to see, which we attribute to the star being relatively faint for tomographic analysis, at $V$ = 11.7, and the transit dip being relatively shallow for a hot Jupiter, at 0.6\%. The latter results from the star being relatively large at 1.6\,R$_{\odot}$ when compared to the planet, which has only a mildly inflated radius of 1.15\,R$_{\rm Jup}$.

\section{Discussion and Conclusions}
\label{sec:discussion}
We have shown that \widb\ is a typical hot Jupiter with a mass of 1.0\,$\pm$\,0.1\,$M_{\rm Jup}$ and a mildly inflated radius of 1.15\,$\pm$\,0.09\,$R_{\rm Jup}$. It is  in a 5.4-day orbit that is marginally misaligned with respect to the stellar rotation, with $\lambda$\,=\,21$\pm$6\,$^{\circ}$.

The measured values of \vsini\ and $\lambda$ are consistent between the spectral analysis, the tomographic analysis and the RM analysis.  The tomographic analysis produced similar fits, giving a \vsini\ value  consistent with the spectroscopic value, regardless of whether we adopted the spectroscopic \vsini\ as a prior. In contrast, the RM analysis was less constrained without a prior, and the fit tended to favour values that were too large. This often occurs for systems with a low impact parameter $b$, since it is difficult to differentiate the effects of \vsini\ and $\lambda$ on the shape of the RM curve when it is symmetrical \citep[e.g. ][]{2011ApJ...738...50A}. Since, in \wid, the impact parameter has a mid-level value of $b$\,=\,0.45, this tendency should be reduced, but it may be that the low signal-to-noise of the data is leading the fit to be less constrained than usual.   Overall, we found that the parameters were better constrained in the tomographic analysis than in the RM analysis, and so we adopt that fit. 

\begin{figure}
\hspace*{2mm}\includegraphics[width=0.5\textwidth]{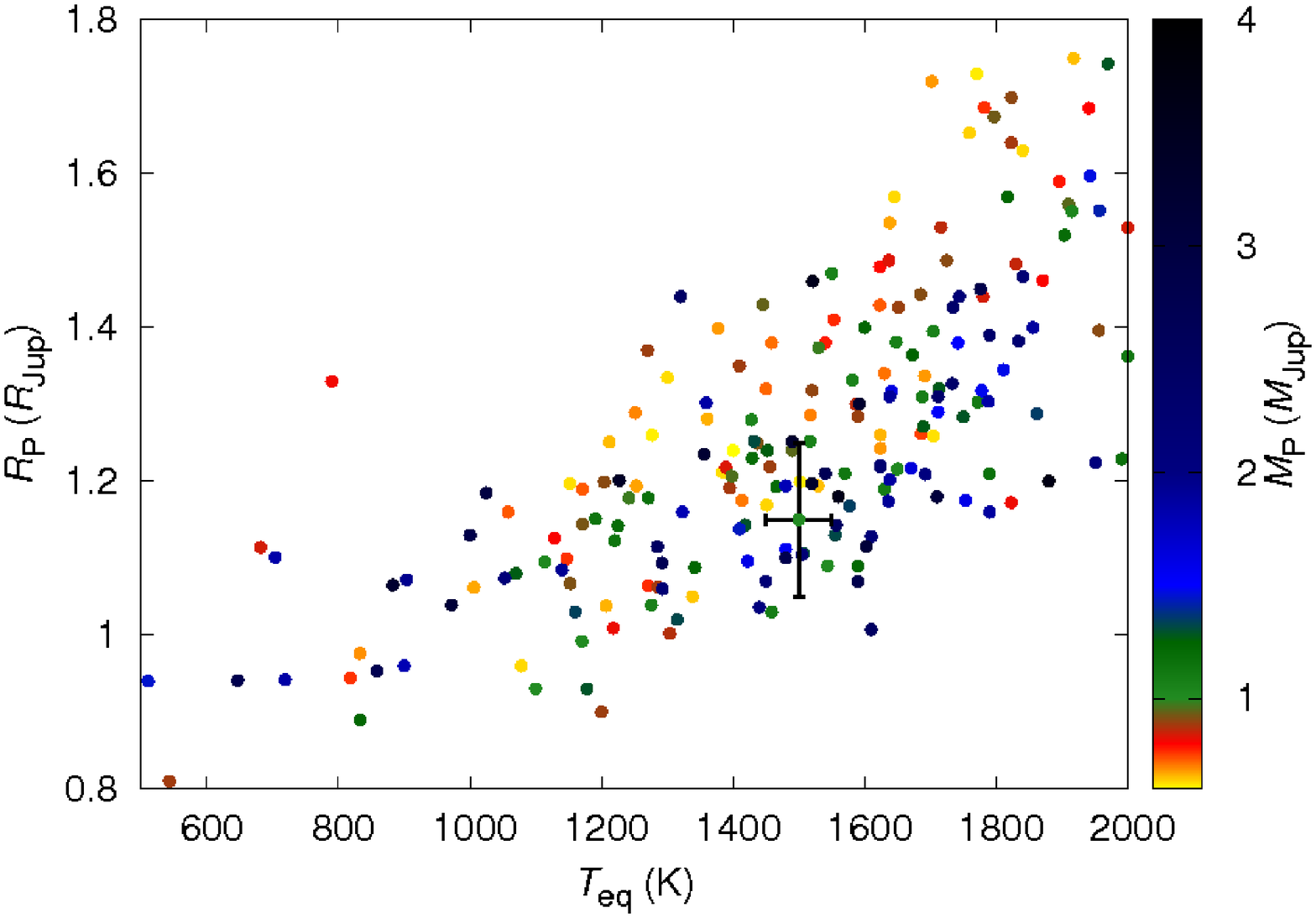}\\ [-2mm]
\centering
\caption{$R_{\rm p}$ vs. $T_{\rm eql}$, colour coded by mass, of all known planets with 0.6\,$M_{\rm Jup}$\,<\,$M_{\rm p}$\,<\,4.0\,$M_{\rm Jup}$. \widb\ is displayed including the error bars on the measured radius and temperature.}
\label{fig:RpvsTeql}
\end{figure}

While there is a well-established trend between the irradiation of a hot Jupiter and the inflation of its radius \citep[e.g.][]{2012A&A...540A..99E}, hot Jupiters also display a wide range of radii \citep[e.g.][]{2007ApJ...661..502B}.  \citet{2018A&A...616A..76S} investigates the relationship between planet radius, mass and irradiation, finding that a more massive planet is usually less inflated than a low-mass planet of the same temperature, due to the planet's gravity counteracting the inflation. In Fig.~\ref{fig:RpvsTeql} we show planetary radius as a function of equilibrium temperature, and use planetary mass as a third dimension, for all planets with 0.6\,$M_{\rm Jup}$\,<\,$M_{\rm p}$\,<\,4.0\,$M_{\rm Jup}$ as listed in the TEPCat database \citep{2011MNRAS.417.2166S}. The figure indicates that planets of a given mass and equilibrium temperature can have a wide range of radii, and shows that planets of $\sim$\,1\,$M_{\rm Jup}$ like \widb\ are not necessarily inflated, implying that the invocation of some third parameter is required.  Possible causes of the disparity include different evolutionary histories, leading to different amounts of irradiation over time \citep[e.g. ][]{2016AJ....152..182H}, the possibility of internal heating mechanisms \cite[e.g.][]{2015ApJ...815...78K,2015ApJ...803..111G, 2018AJ....155..214T,2018MNRAS.481.5517R} and differences in the mass and metallicity of the planets' cores \citep[e.g. ][]{2012A&A...540A..99E}. 

With $\lambda$\,=\,21\,$\pm$\,6$^{\circ}$, \widb\ is marginally misaligned. This is consistent with the known trend in hot-star systems, whereby planets around stars beyond the Kraft break have a wider range of obliquities, with most being in misaligned orbits \citep[e.g.][]{2010ApJ...718L.145W,2017AJ....153..205D}. The true orbit may be more strongly misaligned, however, since the value of $\lvert\lambda\rvert$ for non-polar misaligned orbits represents a lower limit for the true obliquity $\lvert\psi\rvert$. To measure $\psi$ it would be necessary to independently measure the stellar equatorial rotational velocity $v$ or stellar inclination $i_{\star}$ \citep[for example, by looking for differential rotation effects as described by][]{2016A&A...588A.127C}.

\acknowledgments
WASP-South is hosted by the South African Astronomical Observatory and we are grateful for their ongoing support and assistance. Funding for WASP comes from consortium universities and from the UK's Science and Technology Facilities Council. The research leading to these results has received funding from the European Research Council (ERC) under the FP/2007-2013 ERC grant agreement no. 336480, and under the H2020 ERC grant agreement no. 679030; and from an Actions de Recherche Concert\'ee (ARC) grant, financed by the Wallonia-Brussels Federation. The Euler Swiss telescope is supported by the Swiss National Science Foundation (SNF). TRAPPIST-South is funded by the Belgian Fund for Scientific Research (Fond National de la Recherche Scientifique, FNRS) under the grant FRFC 2.5.594.09.F, with the participation of the SNF. M. Gillon and E. Jehin are F.R.S.-FNRS Senior Research Associates. We acknowledge use of the ESO 3.6-m/HARPS spectrograph under programs 0100.C-0847(A), PI C. Hellier, and 0102.C-0414, PI L. Nielsen. This work has made use of data from the European Space Agency (ESA) mission {\it Gaia} (\url{https://www.cosmos.esa.int/gaia}), processed by the {\it Gaia} Data Processing and Analysis Consortium (DPAC, \url{https://www.cosmos.esa.int/web/gaia/dpac/consortium}). Funding for the DPAC has been provided by national institutions, in particular the institutions participating in the {\it Gaia} Multilateral Agreement.

\vspace{5mm}
\facilities{SuperWASP,ESO:3.6m(HARPS), Euler1.2m(CORALIE), WASP-South, TRAPPIST, SPECULOOS}

\software{HARPS Data Reduction Software \citep{1996A&AS..119..373B,2002Msngr.110....9P}, {\sc bagemass} \citep{2015A&A...575A..36M}, {\sc garstec} \citep{2008Ap&SS.316...99W}, MKCLASS \citep{2014AJ....147...80G}}

\bibliographystyle{aasjournal}
\bibliography{litbiblio}

\end{document}